\documentclass[twocolumn,showpacs,amsmath,amssymb]{revtex4}
\usepackage[dvips]{graphics}
\def\imo{i}
\def\re#1{Re(#1)}
\def\im#1{Im(#1)}

\begin{document}
\title{AdS-like spectrum of the asymptotically G\"{o}del space-times}
\author{R. A. Konoplya}\email{konoplya_roma@yahoo.com}
\affiliation{Theoretical Astrophysics, Eberhard-Karls University of T\"{u}bingen, T\"{u}bingen 72076, Germany}
\affiliation{Centro de Estudios Cient\'{i}ficos (CECS), Casilla 1469, Valdivia, Chile.}
\author{A. Zhidenko}\email{olexandr.zhydenko@ufabc.edu.br}
\affiliation{Centro de Matem\'atica, Computa\c{c}\~ao e Cogni\c{c}\~ao, Universidade Federal do ABC,\\ Rua Santa Ad\'elia, 166, 09210-170, Santo Andr\'e, SP, Brazil}
\begin{abstract}
A black hole immersed in a rotating Universe, described by the Gimon-Hashimoto solution, is tested on stability against scalar field perturbations. Unlike the previous studies on perturbations of this solution, which dealt only with the limit of slow Universe rotation $j$, we managed to separate variables in the perturbation equation for the general case of arbitrary rotation. This leads to qualitatively different dynamics of perturbations, because the exact effective potential does not allow for Schwarzschild-like asymptotic of the wave function in the form of purely outgoing waves. The Dirichlet boundary conditions are allowed instead, which result in a totally different spectrum of asymptotically G\"odel black holes: the spectrum of quasinormal frequencies is similar to the one of asymptotically anti-de Sitter black holes. At large and intermediate overtones $N$, the spectrum is equidistant in $N$. In the limit of small black holes, quasinormal modes (QNMs) approach the normal modes of the empty G\"odel space-time. There is no evidence of instability in the found frequencies, which supports the idea that the existence of closed time-like curves (CTCs) and the onset of instability correlate (if at all) not in a straightforward way.
\end{abstract}
\maketitle

\section{Introduction}

Various structures in the universe starting from compact objects, such as stars and planets, and including constellations, galaxies and clusters of galaxies, rotate. This made physicists suppose that the Universe itself can be rotating. The first model for such rotating universe was suggested by K.~G\"odel yet in 1949, who found an exact solution for the $3+1$ dimensional rotating Universe  \cite{Godel} in General Relativity. The solution is homogeneous, has rotational symmetry, and allows for the definition of the direction of positive time consistently in the whole space-time, and, what has been in the focus of recent research. It also allows for closed time-like curves.

In the Einstein gravity a static black hole cannot be embedded in the vacuum G\"odel Universe. However such an embedding is possible if extra fields are added. The asymptotically G\"odel, 5-dimensional, nonrotating black hole solution was obtained by Gimon and Hashimoto \cite{Gimon:2003ms} for the low-energy supergravity limit, i.~e. in the presence of an extra gauge field. This solution attracted considerable interest due to its utility when studying exact solutions in the form of the so-called pp-waves in string theory \cite{string-pp}.

Here we are interested in perturbations of asymptotically G\"odel space-times from a different point of view. First, the G\"odel Universe has a peculiar feature: at its asymptotic, far region it allows for the closed time-like curves (CTCs) which opens possibility of existence of the time machine. Thus, existence of compact objects (for instance, black holes) in such space-times would look most unreal and one could expect that there should be some physical laws which forbid such geometries. Indeed, there are a number of topological theorems which shows that nature should favor ``normal'' causality of the world. Yet, these are rather geometrical arguments, than a rigorous physical reason.

An interesting observation, which might be an example of such a physical reason of the nonexistence objects allowing for CTCs, was made in \cite{Pavan:2009wt}, where it was numerically shown that the infinite cylinder (with an event horizon) in General Relativity has infinitely growing modes in its characteristic spectrum. Thus, the cylindrical space-time, allowing for CTCs, is also gravitationally unstable. From this observation, one could expect that there is a deeper correlation between the onset of the instability and existence of CTCs. A straightforward way is to check other compact objects with CTCs as to their stability. This can be done by investigation of the proper oscillation frequencies (called quasinormal modes \cite{QNMreviews,Berti:2009kk,Konoplya:2011qq} for black holes) of the compact object under consideration. These frequencies dominate at intermediate late time in the black hole's response to the external perturbation.

Quasinormal modes of the Gimon-Hashimoto black holes were first studied in \cite{Konoplya:2005sy}, though the wave equation in \cite{Konoplya:2005sy} was obtained for the slow rotation limit. In this limit, the obtained effective potential looks like the one for the 5-dimensional Schwarzschild black hole with a shift in frequency, which depends on the Universe rotation parameter $j$. In this approximation properties of the asymptotics of the wave equation and, consequently, boundary conditions, are qualitatively the same as for the Schwarzschild black hole. Therefore, quasinormal spectrum is just shifted Schwarzschild spectrum \cite{Konoplya:2005sy}. Following the approximation used in \cite{Konoplya:2005sy}, there appeared a number of consequent works on quasinormal modes and Hawking radiation of more general solutions and other fields \cite{extention}.

Thus, we have a few motivations to study QN spectrum of the asymptotically G\"odel black holes: First is the revealing of possible correlation
between the gravitational instability and existence of CTCs, second, quest for a qualitative insight of main features of the quasinormal spectrum in the rotating Universe, and third, possible utility of QNMs of the exact sting theory background given by the Gimon-Hashimoto solution. Having in mind the above motivations, in the present work we managed to separate angular variables without any approximation through using different coordinate system. The exact wave equation which we obtained revealed qualitatively new feature, which is connected with different asymptotics of the wave equation in the region far from a black hole. We have found that no Schwarzschild-like boundary conditions, that is, purely outgoing waves, are allowed at infinity. Instead Dirichlet boundary conditions should be imposed, in a similar fashion with the asymptotically anti-de Sitter black holes, whose QNMs were extensively studied in the context of gauge/gravity duality (see \cite{Berti:2009kk} and \cite{Konoplya:2011qq} for recent reviews of QNMs).

The paper is organized as follows. In Sec. II we mention some basic properties of the Gimon-Hashimoto solution. Sec III describes separation of variables in the wave equation for a scalar field. Sec. IV is devoted to boundary conditions for the quasinormal modes, while in Sec. V we briefly relate the numerical methods which were used for calculations of the quasinormal modes. Sec. VI discuss particular case of empty G\"odel space-time (without a black hole). Sec VII and VIII are devoted to spectrum of G\"odel black holes, which includes the case of the quasiextremal Universe rotation for which some analytical results were obtained. Finally, we discuss the obtained results and their meaning for stability of the black holes and its possible correlation with the existence of closed time-like curves.

\section{5-dimensional asymptotically G\"odel black hole: the Gimon-Hashimoto solution}

The bosonic fields of the minimal (4+1)- supergravity theory consist of the metric and the one-form gauge field, which are governed by the following equations of motion
\begin{eqnarray}\label{EinsteinEq}
R_{\mu \nu} &=& 2 \left(F_{\mu \alpha} F_{\nu}^{\alpha} -\frac{1} {6} g_{\mu \nu}
F^{2}\right)\,;\\\label{MaxwellEq}
D_{\mu} F^{\mu \nu} &=& \frac{1}{2 \sqrt{3}} \varepsilon^{\alpha \lambda
\gamma \mu \nu}  F_{\alpha \lambda} F_{\gamma \mu}\,;
\end{eqnarray}
Here, $\varepsilon_{\alpha \lambda \gamma \mu \nu} = \sqrt{-\det g_{\mu\nu}}~\epsilon_{\alpha \lambda \gamma \mu \nu}$.

In the Euler coordinates $(t, r, \theta, \psi, \phi)$, the solution of the equations of motion (\ref{EinsteinEq}-\ref{MaxwellEq}), describing the G\"{o}del universe, has the form \cite{Gimon:2003ms}:
\begin{eqnarray}
ds^2 &=& - (dt + jr^2 \sigma_{L}^{3})^2 + dr^{2} \nonumber\\
&&+\frac{r^2}{4}(d \theta^{2} + d \psi^{2} + d \phi^{2} + 2 \cos \theta d \psi d \phi),
\end{eqnarray}
where $\sigma_{L}^{3}= d \phi + cos \theta d \psi$. The parameter $j$
defines the scale of the G\"{o}del background.  At $j=0$ we have
the Minkowski space-time. The solution for the  Schwarzschild black hole
in the G\"{o}del universe is given by \cite{Gimon:2003ms}
\begin{eqnarray}
ds^2 &=& - f(r) dt^2 -g(r) r \sigma_{L}^{3} d t - h(r) r^2
(\sigma_{L}^{3})^2 + k(r) d r^2 \nonumber\\
&&+\frac{r^2}{4}(d \theta^{2} + d \psi^{2} + d \phi^{2} + 2 \cos \theta d \psi d \phi),\label{metric}
\end{eqnarray}
where
\begin{eqnarray}\label{coeff}
g(r) = 2 j r,& \quad &h(r)=j^2 (r^2 + 2 M),\\\nonumber
f(r)= 1- \frac{2 M}{r^2},&  \quad &k(r) = \left(1 - \frac{2 M}{r^2} + \frac{16 j^2 M^2}{r^2}\right)^{-1}.
\end{eqnarray}
The radius of the event horizon is also corrected by parameter $j$,
\begin{equation}
r_0 = \sqrt{2 M (1- 8 j^2 M)}.
\end{equation}

An essential feature of the above Gimon-Hashimoto solutions is possibility of the closed time-like curves (CTCs) in a
region far from the black hole,
\begin{equation}
r > \frac{\sqrt{1- 8 M j^2}}{2 j}.
\end{equation}

\section{Separation of the variables}

In order to derive the wave equation we only need the relation
$$g^2(r)+f(r)(1-4h(r))=\frac{1}{k(r)}\,,$$
which implies
$$det g_{\mu \nu}=-\frac{r^6\sin^2\theta}{64}\,.$$

The nonvanishing components of the inverse metric have the following form
\begin{eqnarray}
\nonumber&& g^{tt} = -(1-4 h(r))k(r),
\quad g^{rr} = k^{-1}(r),
\quad g^{\theta \theta} = 4/r^{2},\\
\nonumber&& g^{\psi \psi} = \frac{4}{r^2 \sin^{2} \theta},
\quad g^{\phi \phi} = \frac{4 (\cot^2\theta+k(r)f(r))}{r^2},\\
\nonumber&& g^{t \phi} = - \frac{2 g(r) k(r)}{r},
\quad g^{\theta \psi} = - \frac{4 \cos \theta }{r^2 \sin^{2} \theta}.
\end{eqnarray}

The scalar field perturbations in a curved background are governed by
the Klein-Gordon equation
\begin{equation}\label{Klein-Gordon}
\Box \Phi  \equiv \frac{1}{\sqrt{-g}} \partial_\mu\left(g^{ \mu \nu} \sqrt{-g}
\partial_\nu\Phi\right)  = \mu^2\Phi.
\end{equation}
Since the background metric has the Killing vectors  $\partial_{t}$,
$\partial_{\psi}$,  $\partial_{\phi}$, we choose the ansatz for the wave function as
\begin{equation}\label{ansatz}
\Phi(t,r,\theta,\psi,\phi) = e^{\displaystyle-\imo \omega t + \imo n \psi + \imo m \phi} Y(\theta) R(r)r^{-3/2}.
\end{equation}

Substituting (\ref{ansatz}) into (\ref{Klein-Gordon}) and separating the variables, one can find that the angular part of the function satisfies the equation
\begin{equation}\label{angular}
\left(\frac{1}{\sin\theta}\frac{d}{d\theta}\sin\theta\frac{d}{d\theta}+\frac{2mn\cos\theta-m^2-n^2}{\sin^2\theta}+\lambda\right)Y(\theta)=0,
\end{equation}
where $\lambda$ is the separation constant with the eigenvalues
$$\lambda=\ell(\ell+1), \quad \ell=\max(|m|,|n|)+i, \quad i=0,1,2\ldots.$$

The equation for the radial part takes the wave-like form
\begin{equation}\label{wave-like}
\left(\frac{d^2}{dr_\star^2}+Q(r_\star)\right)R(r_\star)=0,
\end{equation}
where $r_\star$ is the tortoise coordinate, which is defined as
\begin{equation}\label{tortoise}
dr_\star=k(r)dr.
\end{equation}
The effective potential can be written in terms of the coordinate $r$ as follows
\begin{eqnarray}
&&Q(r)=\left(1-4h(r)\right)\left(\omega-\frac{2mg(r)}{r(1-4h(r))}\right)^2-
\\\nonumber&&\frac{1}{k(r)}\left(\frac{4\lambda}{r^2}+\mu^2+\frac{16m^2h(r)}{(1-4h(r))r^2}+\frac{3k(r)-6rk'(r)}{4r^2k^2(r)}\right)\,.
\end{eqnarray}

Now we are in position to perform numerical analysis of the quasinormal spectrum for the obtained wave equation.

\section{Boundary conditions}

At the classical level, a wave cannot be emitted from the black hole horizon, so that the standard
boundary condition for a great variety of black holes is requirement of the purely ingoing wave at the event horizon,
\begin{equation}\label{hor-ingoing}
R\propto e^{-\imo\tilde{\omega} r_\star\sqrt{1-4h(r_\star)}}, \qquad r_\star\rightarrow-\infty,
\end{equation}
where
$$\tilde{\omega}=\omega-\frac{2mg(r_0)}{r_0-4r_0h(r_0)}.$$

Taking account of $\left(k^{-1}(r_0)\right)'=2/r_0$, we find that
\begin{equation}\label{hor-asymptot}
R\propto \left(r-r_0\right)^{-\displaystyle\imo\tilde{\omega}r_0\sqrt{1-4h(r_0)}/2}, \qquad r\rightarrow r_0.
\end{equation}

At spatial infinity we have
\begin{equation}\label{inf-asymptot}
R(r\rightarrow\infty)=C_+\Psi_+(r)+C_-\Psi_-(r),
\end{equation}
where
$$\Psi_\pm(r)=e^{\pm j\omega r^2}r^{\alpha_\pm}\left(1+\frac{A_{1\pm}}{r}+\frac{A_{2\pm}}{r^2}+\frac{A_{3\pm}}{r^3}\ldots\right),$$
with
$$\alpha_\pm=-\frac{1}{2}\pm\left(2m+6\omega j M - 32\omega j^3 M^2-\frac{\omega^2-\mu^2}{4\omega j}\right).$$

Since the exponents $e^{\pm j\omega r^2}$ have purely real index, they do not describe ingoing or outgoing waves. Therefore, we are unable to impose usual quasinormal boundary conditions. However, we can use the analogy with AdS backgrounds and require Dirichlet boundary conditions at spatial infinity. This implies that $C_+=0$ for $\re{j\omega}>0$ or $C_-=0$ for $\re{j\omega}<0$.

The nontrivial behavior of the functions $\Psi_\pm$ is observed when $\re{\omega}=0$. In this case both exponents have oscillatory behavior at spatial infinity. Thus, in order to impose the Dirichlet boundary conditions we must consider the factor $r^{\alpha_\pm}$. When $\re{\omega}=0$ one can find that $$\re{\alpha_\pm}=-\frac{1}{2}\pm2m,$$
implying that one of $\Psi_\pm$ is convergent and the other one is divergent as $r\rightarrow\infty$. The only exception is $m=0$, when both $\Psi_+$ and $\Psi_-$ are convergent.

For this case we recall that we require the function norm be convergent at spatial infinity, i.~e., if $R\propto1/\sqrt{r}$, then the norm
\begin{equation}\label{divergent}
\int|\Phi|^2\sqrt{-g}d^4x\propto\int|R|^2dr\propto\int\frac{dr}{r}\sim \ln r
\end{equation}
is divergent. The asymptotically anti-de Sitter space-times allow for a family of boundary conditions and a family of norms. For lower dimensional AdS black holes, such as the 2+1 dimensional BTZ black holes \cite{Banados:1992wn}, the vanishing flux at infinity is used sometimes
as the boundary condition \cite{Cardoso:2001hn,Konoplya:2004ik}. If instead of the above norm one uses the flux of the field, such norm will be even stronger divergent for the $m=0$ case.

\begin{figure*}
\resizebox{\linewidth}{!}{\includegraphics*{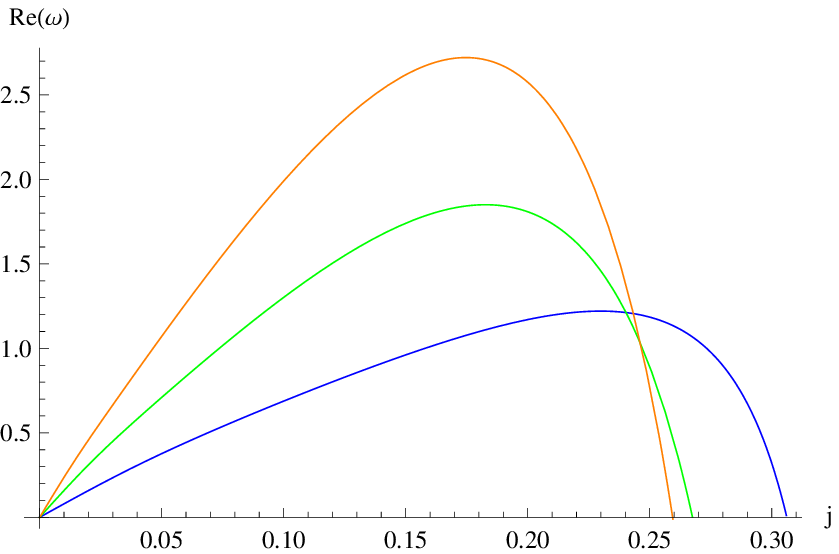}\includegraphics*{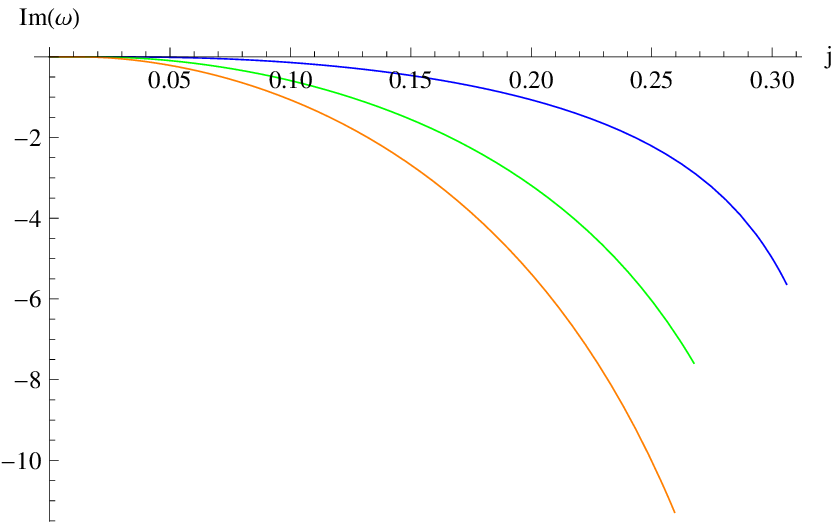}}
\caption{Real and imaginary parts of first three dominant quasinormal frequencies as functions of $j$ ($M=1$, $\mu=0$) for $\lambda=0$. The real part of the dominant frequency is smaller and approaches zero for the largest threshold value of $j$. Higher overtones have larger oscillation frequency and damping rate.}\label{fig:fundamental}
\end{figure*}

Thus, the boundary conditions do not allow us to have purely imaginary quasinormal mode for $m=0$. However, we observe that $\re{\omega}$ approaches zero at some value of the rotation parameter (see Fig. \ref{fig:fundamental}). Since we cannot find any quasinormal mode for larger values of the rotation parameter, probably, one should interpret this as disappearing of the particular mode after this threshold value of $j$ is reached.

\begin{figure*}
\resizebox{\linewidth}{!}{\includegraphics*{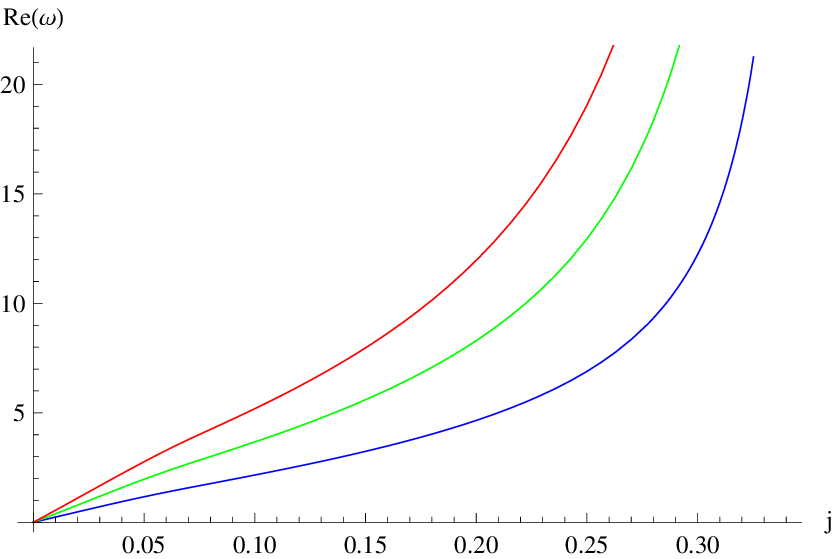}\includegraphics*{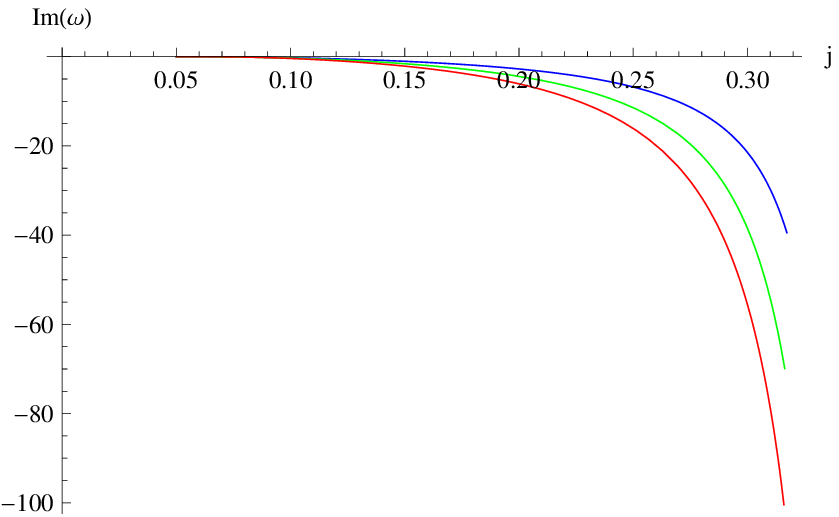}}
\caption{Fundamental modes as functions of $j$ for $\ell=m=1$ (blue, bottom), $\ell=m=2$ (green, middle), $\ell=m=3$ (red, top) ($M=1$, $\mu=0$).}\label{fig:m=123}
\end{figure*}

\begin{figure*}
\resizebox{\linewidth}{!}{\includegraphics*{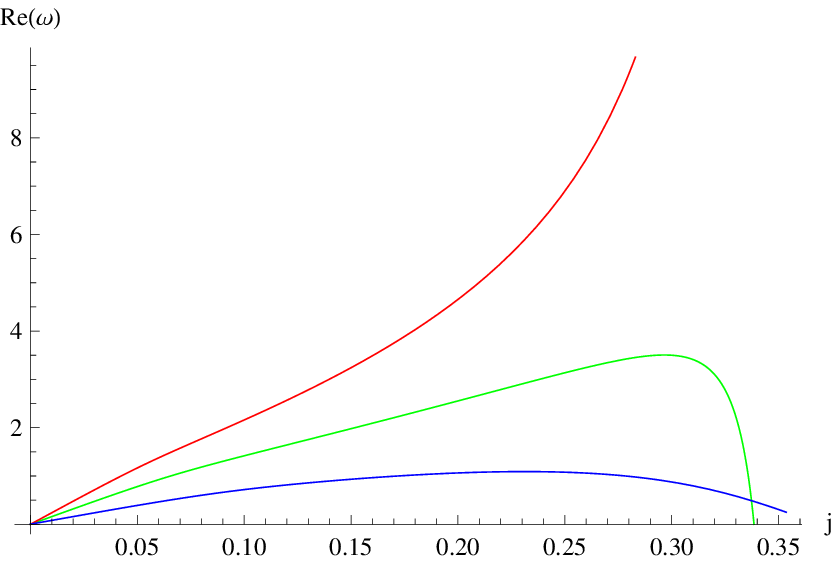}\includegraphics*{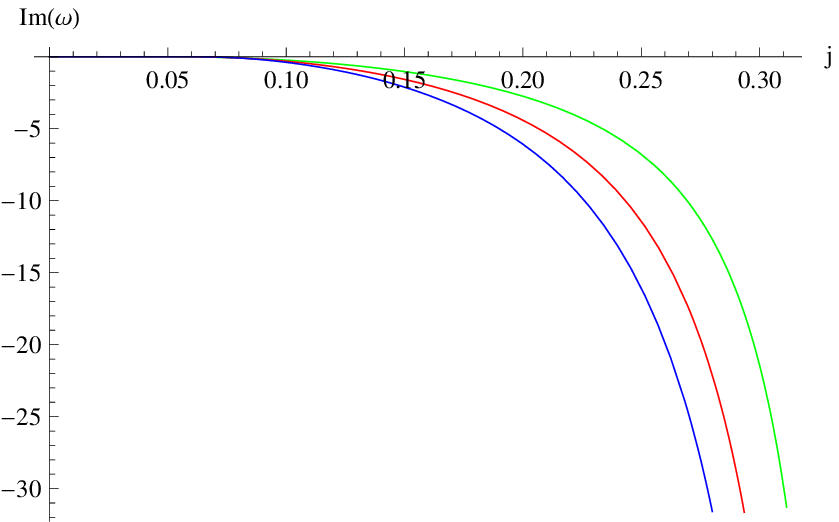}}
\caption{Fundamental modes as functions of $j$ for $\ell=1$  $m=-1$ (blue, bottom), $m=0$ (green, top on the left pane), $m=1$ (red, top on the right pane) ($M=1$, $\mu=0$).}\label{fig:l=1}
\end{figure*}

\begin{figure*}
\resizebox{\linewidth}{!}{\includegraphics*{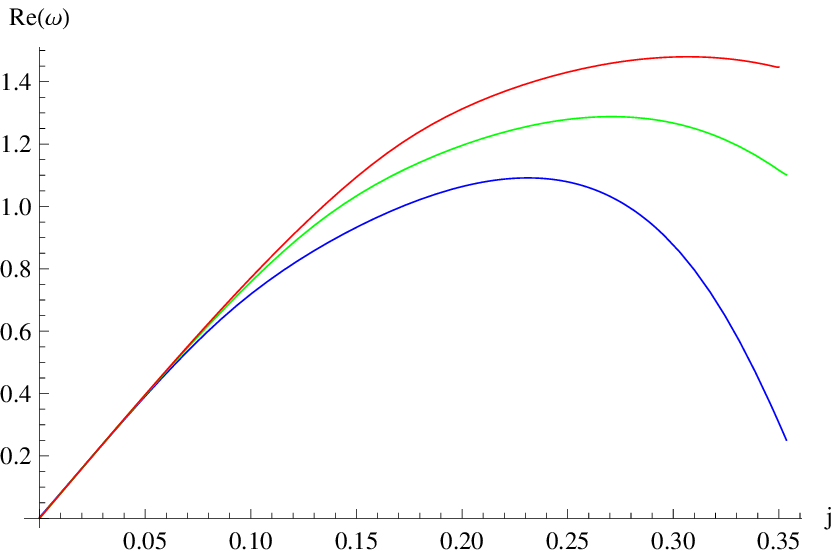}\includegraphics*{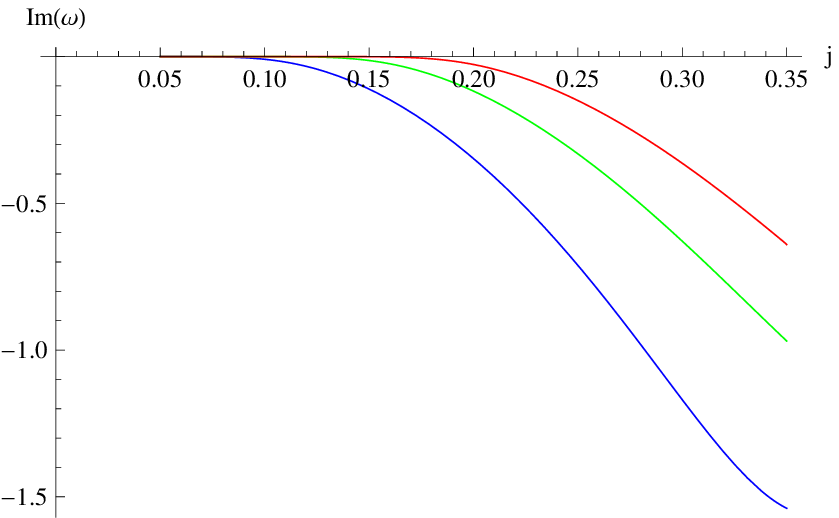}}
\caption{Fundamental modes as functions of $j$ for $\ell=-m=1$ (blue, bottom), $\ell=-m=2$ (green, middle), $\ell=-m=3$ (red, top) ($M=1$, $\mu=0$).}\label{fig:m=-123}
\end{figure*}

We observe this ``disappearing'' of the quasinormal modes only for the case of $m=0$. The quasinormal modes with the positive real frequency exist for all values of $j$, if $m>0$ (see Fig. \ref{fig:m=123}). For negative $m$ we observe the same tendency to ``disappear'', though for the fundamental mode the extremal value of $j$ is reached before the real part vanishes (see Figs. \ref{fig:l=1}, \ref{fig:m=-123}). Since for higher overtones, modes disappear at lower threshold values of $j$, we may expect that for some near-extremal value of the rotation parameter we have only finite number of the modes with positive real part. Because of the symmetry of the spectrum,
\begin{equation}\label{specsym}
m\rightarrow-m,\qquad \re{\omega}\rightarrow-\re{\omega},
\end{equation}
we still have infinite number of modes with negative value of the real part.

\section{Numerical methods}

Here, we shall briefly relate two standard methods used for numerical search of quasinormal modes: the shooting method and Frobenius method. These two alternative methods were used in order to guarantee the validity of the obtained results. More detailed discussion of both methods can be found in \cite{Konoplya:2011qq}.

\textbf{Shooting method}. Since at the horizon we require the purely ingoing wave (\ref{hor-asymptot}), it is convenient to define a new function in such a way,
\begin{equation}\label{reg-function}
y(r)=\left(1-\frac{r_0^2}{r^2}\right)^{\displaystyle\imo\tilde{\omega}r_0\sqrt{1-4h(r_0)}/2}\times R(r),
\end{equation}
that it becomes regular at the event horizon.

We fix the wave-function norm so that $y(r_0)=1$. Substituting (\ref{reg-function}) into (\ref{wave-like}) and expanding the wave equation near the horizon, we find that $y'(r_0)$. This gives us boundary condition at the horizon for any fixed $\omega$, which we use for the numerical integration of the equation (\ref{wave-like}). At large distances we compare the result of our numerical integration with the large-distance asymptotic series expansion (\ref{inf-asymptot}) and find the coefficients $C_\pm$ by using the fitting procedure. Then, quasinormal modes can be found by minimizing $C_+(\omega)$ or $C_-(\omega)$ depending on the sign of $\re{\omega}$. In order to check convergence of the procedure we check that obtained frequencies do not change within specified accuracy, if we increase precision of floating-point operations or the distance at which we use the fit.

\textbf{Frobenius method.} This method allows us to find QNMs by solving numerically an equation with continued fractions, which takes much less computer time for finding quasinormal modes. One can rewrite the wave-function as
\begin{equation}\label{frobenius-prefactor}
R(r)=e^{\mp \omega jr^2}r^{\alpha_\mp}\left(1-\frac{r_0^2}{r^2}\right)^{-\displaystyle\imo\tilde{\omega}r_0\sqrt{1-4h(r_0)}/2}\times z(r),
\end{equation}
where $z(r)$ must be regular at spatial infinity and the event horizon, once $\omega$ is the quasinormal frequency.
We choose ``-'' sign for $\re{\omega j}>0$, and ``+'' sign for $\re{\omega j}<0$.

The function $z(r)$ can be expanded into a series near the horizon
\begin{equation}\label{frobenius}
z(r)=\sum_{i=0}^{\infty}a_i\left(1-\frac{r_0^2}{r^2}\right)^i.
\end{equation}
After we substitute (\ref{frobenius-prefactor}) and (\ref{frobenius}) into equation (\ref{wave-like}), we find the three-term recurrence relation for the coefficients $a_i$
\begin{eqnarray}\label{three-terms}
&&c_{0,i}\,a_i+c_{1,i}\,a_{i-1}+c_{2,i}\,a_{i-2}=0, \quad
\,i>1\\
&&c_{0,1}\,a_1+c_{1,1}\,a_0=0.
\end{eqnarray}

From the above, the coefficients of the three-term recurrence relation can be found in a closed form. Finally, we find the equation with the infinite continued fraction on the righthand side
\begin{eqnarray}
c_{1,N+1}-\frac{c_{2,N}c_{0,N-1}}{c_{1,N-1}-}
\,\frac{c_{2,N-1}c_{0,N-2}}{c_{1,N-2}-}\ldots\,
\frac{c_{2,2}c_{0,1}}{c_{1,1}}=\nonumber\\\frac{c_{0,N+1}c_{2,N+2}}{c_{1,N+2}-}
\frac{c_{0,N+2}c_{2,N+3}}{c_{1,N+3}-}\ldots\,,\label{invcf}
\end{eqnarray}
which we solve numerically with respect to the quasinormal frequencies $\omega_N$ \cite{Leaver:1985ax}.

\section{The limit of pure G\"odel $D=5$ space-time (without a black hole)}\label{sec:nobh}

Let us take the limit $M \rightarrow 0$ in the wave equation (\ref{wave-like}). Then, $k(r) = f(r) =1$, the tortoise coordinate $r_\star = r$  and the wave equation (\ref{wave-like}) is reduced to a much simpler form
\begin{equation}\label{nobheq}
\frac{d^2 R}{dr^2} + \left((1 - 4 j^2 r^2)\omega^2 - 8 j m \omega- \frac{\frac{3}{4} + 4 \lambda}{r^2} - \mu^2)\right) R =0.
\end{equation}

Substituting $\lambda=\ell(\ell+1)$ and
$$R=\left\{\begin{array}{ll}e^{-j\omega r^2}r^{-2\ell-1/2}p(r), & \re{j\omega}>0, \\e^{j\omega r^2}r^{-2\ell-1/2}p(r), & \re{j\omega}<0,\end{array}\right.
$$
we obtain the Kummer's equation
\begin{equation}
zp''(z)+(b-z)p'(z)=ap(z),
\end{equation}
with respect to the new coordinate $$z=\left\{\begin{array}{ll}2j \omega r^2, & \re{j\omega}>0, \\-2j \omega r^2, & \re{j\omega}<0.\end{array}\right.$$
Here $b=-2\ell$ and
\begin{equation}\label{cond}8j\omega a=\left\{\begin{array}{ll}-8j(\ell-m)\omega-\omega^2+\mu^2, & \re{j\omega}>0, \\-8j(\ell+m)\omega+\omega^2-\mu^2, & \re{j\omega}<0.\end{array}\right.\end{equation}
The solution of the Kummer's equation, which produces regular at the origin ($r=0$) wave-function $\Psi(r)$, is
$$p(z)\propto z^{1+2\ell} M(a+2\ell+1,2\ell+2,z).$$
Here $M(A,B,z)$ is the generalized hypergeometric series, which has an irregular singularity at $z=\infty$, unless $A$ is a nonpositive integer. Thus, one has
$$a=-N-2\ell, \qquad N=1,2,3\ldots .$$
Finally, from (\ref{cond}) we find
\begin{equation}\label{Godelqnms}
\omega=\pm4j\left(N+\ell\pm m+\sqrt{(N+\ell\pm m)^2+\frac{\mu^2}{16j^2}}\right).
\end{equation}

One can see an interesting features of the scalar waves in $D=5$ G\"odel Universe: there is a minimal value of frequency for any multipole number $\ell$, which is $\omega = 8 j$ for $\mu=0$ (see Fig. \ref{fig:smallM}). Waves with lower frequencies cannot propagate in the G\"odel Universe. A similar property was observed by Hiscock for the four-dimensional G\"odel Universe in \cite{Hiscock}.

There are both positive and negative frequencies in the spectrum. Negative ones were interpreted by Hiscock \cite{Hiscock} as those propagating back in time, because of the allowed closed time-like curves. Indeed, let us associate a wave with a negative frequency as that propagating back in time. Then, due to the symmetry (\ref{specsym}), we can conclude that such a wave with a given frequency corresponds to the opposite value of the azimuthal number $m$, which is eigenvalue of the time-like Killing vector in the region with CTCs. This means that the wave, propagating back in time in the region near the black hole, also propagates back in time when it reaches the region with CTCs.

We can see that the perturbation spectrum of the five-dimensional G\"odel space-time is qualitatively similar to the four-dimensional spectrum \cite{Hiscock}, and resembles the spectrum of empty anti-de Sitter space-time \cite{Konoplya:2002zu,Burgess:1984ti}, as both consist of equally spaced normal modes.

\section{Spectrum of a black hole in the G\"odel Universe}

In general, quasinormal modes $\omega$ are complex and $\re{\omega}$ is the real oscillation frequency, while $\im{\omega}$ is proportional to the damping rate of a given mode. For normal modes of pure G\"odel space-times discussed above $\im{\omega} =0$, so that waves look like nondamping (standing) waves in a confining box. Another essential property of quasinormal modes is their significance for stability analysis: if all QN modes are damped, the black hole is believed to be stable, while existence of any growing mode indicates gravitational instability. Analytically stability is guaranteed if the effective potential is positive definite outside the black hole, which happens only in a few simplest cases, i.~e. for the Schwarzschild back hole. More general cases, and the G\"odel black hole is one of them, require further numerical approach to the wave equation for the analysis of of stability. Quasinormal modes have been effectively used in a number of works for checking of the gravitational stability (or finding of instability) of various black holes \cite{stability}.

\begin{figure*}
\resizebox{\linewidth}{!}{\includegraphics*{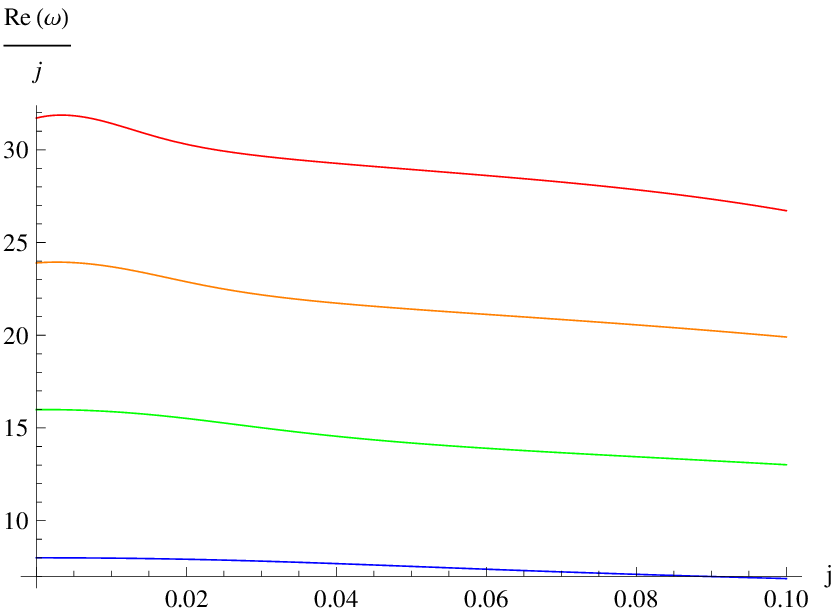}\includegraphics*{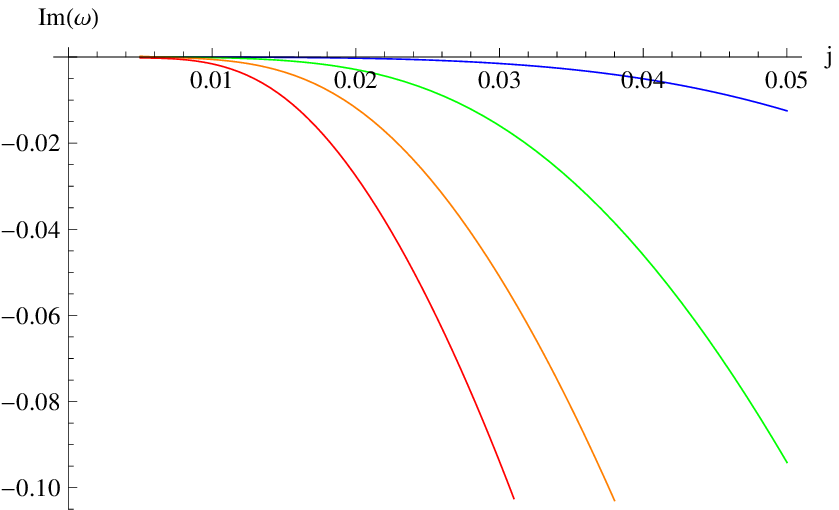}}
\caption{First four dominant quasinormal frequencies as functions of $j$ ($M=1$, $\mu=0$) for $\lambda=0$. As $j$ approaches zero $\re{\omega}\simeq 8Nj$ and $\im{\omega}\propto j^4$. Higher overtones have larger oscillation frequency and damping rate.}\label{fig:smallj}
\end{figure*}

\begin{figure*}
\resizebox{\linewidth}{!}{\includegraphics*{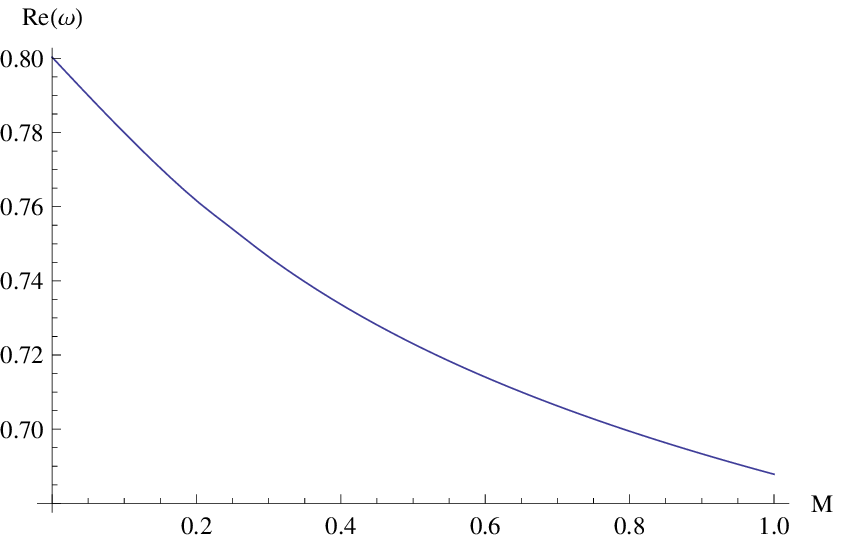}\includegraphics*{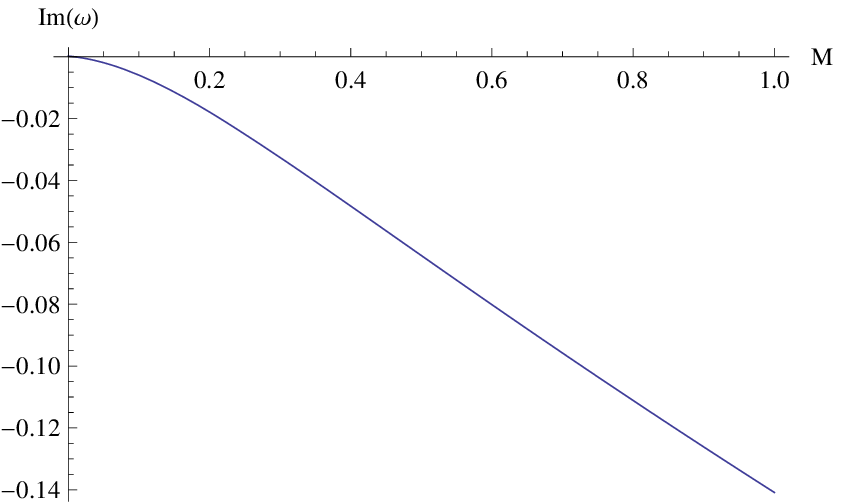}}
\caption{Fundamental mode approaches $8j$ as $M\rightarrow0$ ($j=0.1$, $\lambda=0$, $\mu=0$).}\label{fig:smallM}
\end{figure*}

The pure G\"odel space-time has one scale parameter $j$. This parameter defines a scale of the universe which is distance $\propto j^{-1}$ to the region where the closed time-like geodesic trajectories exist. When we add a black hole to the G\"odel Universe, another scale parameter appears, which is the black hole radius $r_0$. In this paper we consider fixed black hole masses and various universe's rotation parameters $j$. However, instead one could fix the scale of the universe $j$ and look at various $r_0$. In the latter case, black holes are parameterized by the dimensionless quantity $M j^2$. For small values of $j$, $M j^2$ quickly approaches zero and the metric (\ref{metric}) tends to the pure G\"odel space-time with the quasinormal spectrum given by the normal modes of G\"odel space-time (\ref{Godelqnms}), what is confirmed by modes shown on Fig. \ref{fig:smallj}.

In concordance with this, on Fig. \ref{fig:smallM} it is shown that as $M \rightarrow 0$, that is $M j^2 \rightarrow 0$ at a fixed $j$, QNMs approach normal modes of the G\"odel space-time. This behavior exactly resembles the one of the asymptotically AdS black holes, for which QNMs approach normal modes of AdS space-time in the limit $r_0 \rightarrow 0$ \cite{Konoplya:2002zu}. As $j$ approaches zero at a fixed $M$, we have the limit of normal modes of the G\"odel space-time
\begin{equation}
\re{\omega}\simeq 8Nj, \quad  \im{\omega}\propto j^4, \quad j \rightarrow 0.
\end{equation}

Thus, at small $j$ QNMs of black holes in the G\"odel Universe are shorter lived and have larger real oscillation frequency, when $j$ increases.

The numerical search of modes includes a kind of indeterminacy because a mode is a root of quite complicated and lengthy algebraic equation, so that ad hoc we do not know which mode corresponds to which overtone number $N$. At the same time, we are most interested in fundamental modes $N=1$, which are longest lived and are, therefore, dominate at late time. In order to track value of $N$ of each mode, we started from very small values of $j$ for which the quasinormal modes are close to the normal modes, and increased gradually the rotation parameter $j$.

The Universe naturally contains only very \emph{small relatively the Universe scale} black holes. Therefore, quasinormal modes of asymptotically G\"odel black holes are inevitably very close to the normal modes of the G\"odel Universe. In other words, the quasinormal spectrum which we obtained represents rather the ringing of the Universe itself slightly corrected by the presence of a black hole. A somewhat more phenomenological approach to the black hole in the rotating Universe was suggested in \cite{Konoplya:2005sy}, in the limit of small universe's rotation. In \cite{Konoplya:2005sy} the asymptotic of the wave equation is qualitatively the same as in the 5-dimensional Schwarzschild case and the spectrum is Schwarzschild-like, though with some corrections proportional to $j$.

Another interesting phenomena can be noticed on Fig. \ref{fig:fundamental}: $\re{\omega}$ vanishes after reaching some critical, moderate value of $j$, so that the mode cannot be numerically detected at larger $j$. As we could not fully investigate the limit of quasiextremal rotation $j$ numerically, below we shall try to analyze the wave equation analytically in this limit.

QNMs which are shown on Fig. 1 - 6, although obtained for small and moderate $j$, do not demonstrate any tendency to instability, because the damping rate, which is proportional to $ Im(\omega)$, increases as $j$ grows. We have not reached, however, the extremal values of $j$ for all modes and, therefore, cannot state that there is no instability for all $j$.

\section{The limit of the quasiextremal black hole in the G\"odel Universe}

In the near-extremal black hole limit one has $M \rightarrow 1/8j^2$, while the event horizon approaches zero. Implying that this situation is rather exotic, let us consider a universe rotating with speed very close to its extremal one, so that the radius of a black hole will still be finite and very close to zero.
In this case, $k(r) \rightarrow 1$ and the tortoise coordinate $r_\star$ again coincides with $r$.
Then, the wave equation takes the following form
\begin{equation}\label{extremaleq}
\frac{d^2 R}{dr^2} + \left(\frac{m^2}{j^2r^4} - 4 j^2 r^2\omega^2 - 8 j m \omega- \frac{\frac{3}{4} + 4 \lambda}{r^2} - \mu^2\right) R =0.
\end{equation}

When $m \neq 0$ equation (\ref{extremaleq}) has two irregular singular points: $r=0$ and $r=\infty$, but when $m=0$, $r=0$ becomes a regular singular point and the general exact solution can be found. For the massless field ($\mu =0$) it is given by
\begin{equation}
R = C_1 \sqrt{r} I_{-\ell-1/2}(\omega jr^2) + C_2 \sqrt{r} I_{\ell+1/2}(\omega jr^2),
\end{equation}
where $I_\alpha(z)$ is the modified Bessel function of the first kind. One of these Bessel functions converges at infinity, satisfying the Dirichlet boundary condition, while the other one diverges. However, this function is divergent also at $r=0$. It is natural to expect that the function is divergent at the horizon. The radius of the horizon approaches zero when considering quasiextremal black holes. In order to have physically meaningful solution one has to require purely ingoing wave at the horizon, which is not possible for the extremal limit. Thus, formally the problem of the adequate boundary conditions in the extremal case remains open. If we require the wave-function to have a finite norm, there are no quasinormal modes allowing it, what is consistent with the numerical results in the near-extremal region.

Since (\ref{extremaleq}) coincides with (\ref{nobheq}) for $m=0$ and $\mu^2\rightarrow\mu^2-\omega^2$ we repeat the procedure of Sec. \ref{sec:nobh} and find that
\begin{equation}8j\omega a=\left\{\begin{array}{ll}-8j\ell\omega+\mu^2, & \re{j\omega}>0, \\-8j\ell\omega-\mu^2, & \re{j\omega}<0,\end{array}\right.\end{equation}
what cannot be satisfied for $\mu^2\geq0$ because $a+\ell <0$.

For the particular case $\ell=0$, $\mu=0$, we can easily find the solution which satisfies the Dirichlet boundary conditions at spatial infinity
$$R(r)\propto e^{\pm j\omega r^2}/\sqrt{r}.$$
However, similarly to (\ref{divergent}), one can prove that this solution leads to a divergent norm as $r\rightarrow0$.

Thus, we conclude that \emph{for $m = 0$ there is no solution}, which satisfies the quasinormal boundary conditions. Numerical calculations show that this takes place in some near-extremal region as well. Physically we may interpret this as nonexistence of the most symmetrical perturbations of the quasiextremely rotating G\"odel Universe, which can still be considered small. In case of arbitrary values of $m$ and
$\ell$ the question of stability of perturbations remains open.

\section{Discussions}

\begin{table}
\caption{Similarities in QN spectrum of asymptotically AdS and G\"odel space-times.}\label{thetable}
\begin{tabular}{|c|c|}
  \hline
  asymptotically AdS BHs & asymptotically G\"odel BHs \\
  AdS radius $R$ & Universe's scale $j^{-1}$ \\
 \hline
  small BH $(r_0\ll R)$ & small BH $(r_0\ll j^{-1})$ \\
  $\approx$ normal modes of AdS & $\approx$ normal modes of G\"odel \\
 \hline
 \multicolumn{2}{|c|}{QNMs are equidistant at large $N$}\\
  \hline
\end{tabular}
\end{table}

In the present work we have studied the scalar field perturbations around the Giomon-Hashimoto solution, which is the generalization of 5-dimensional Schwarzschild solution immersed in the G\"odel Universe. The main result of this paper is that the quasinormal spectrum of such ``Schwarzschild-G\"odel'' black hole is totally different from the Schwarzschild spectrum and resembles the spectrum of asymptotically anti-de Sitter black holes (see Table \ref{thetable}). The latter is due to the Dirichlet boundary conditions at spatial infinity imposed for the asymptotically G\"odel space-times. The numerical analysis shows no signs of instability at small and moderate values of the Universe's scale $j$, which means that the conjectured correlation of the gravitational instability and existence of closed time-like curves, if it exists, is not straightforward.

The frequencies $\omega$ are eigenvalues of the Killing vector $\partial_t$ which is timelike in some region near black hole. In the region far form the black hole we have CTCs and the timelike Killing vector is $\partial_\phi$ with the integer eigenvalue $m$. We cannot have a consistent solution which is not periodic in the region with CTCs. Thus we can think about existence or nonexistence of growing oscillations only in the region near the black hole. The region with CTCs always implies periodical dynamics in it. This periodical behavior of the scalar field in the asymptotical region effectively provides a ``confining box'' from the point of view of the observer near a black hole in the G\"odel Universe. That is why we require the Dirichlet boundary condition at infinity. From the mathematical viewpoint this boundary condition is dictated by the structure of the irregular singularity at $r=\infty$.

In the future we would like to consider perturbations of rotating analog of Gimon-Hashimoto solution, which should show much reacher physics. In particular, we expect that the effective confining box given by the Dirichlet boundary condition together with rotation, and thus with superradiance, should produce the superradiant instability in the regime of quick rotation \cite{KZ-inprogress}.

\begin{acknowledgments}
This work was partially supported by the Alexander von Humboldt foundation, Germany. R. A. K. acknowledges hospitality of the
Centro de Estudios Cient\'{i}ficos (CECS) in Valdivia (Chile). R. A. K. would like also to thank Ricardo Troncoso and Jorge Zanelli
for useful discussions. A.~Z. was supported by Conselho Nacional de Desenvolvimento Cient\'ifico e Tecnol\'ogico (CNPq).
\end{acknowledgments}

\end{document}